\documentclass[times,onecolumn,preprint,authoryear]{elsarticle}
\usepackage{jasr}
\usepackage{framed,multirow}

\usepackage{amssymb}
\usepackage{latexsym}

\usepackage[switch]{lineno}

\usepackage{url}
\usepackage{xcolor}
\definecolor{newcolor}{rgb}{.8,.349,.1}

\usepackage[citebordercolor=white]{hyperref}

\journal{Advances in Space Research}

\begin{document}

\verso{Strauss \textit{etal}}

\begin{frontmatter}

\title{First results from the ENTOTO neutron monitor: Quantifying the waiting time distribution}

\author[NWU]{R.~D.~Strauss, dutoit.strauss@nwu.ac.za}

\author[ESSTI]{Nigussie M.~Giday, nigussiem@essti.gov.et}

\author[ESSTI]{Ephrem B.~Seba, ephremb@essti.gov.et}

\author[ESSTI]{Daniel A.~Chekole, danielat@essti.gov.et}

\author[ESSTI]{Gemechu F.~Garuma, gemechuf@essti.gov.et}

\author[AASTU]{Bereket H.~Kassa, bereketb2j@gmail.com}

\author[ESSTI]{Teshome Dugassa, teshomed@essti.gov.et}

\author[NWU]{C.~Diedericks, 27863840@nwu.ac.za}

\address[NWU]{Center for Space Research, North-West University, Potchefstroom Campus, Private Bag X6001, 2520 Potchefstroom, South Africa}

\address[ESSTI]{Department of Space Science and Applications Research, Ethiopian Space Science and Technology Institute, P.O. Box 33679, Addis Ababa, Ethiopia}

\address[AASTU]{Department of electrical engineering, Addis Ababa Science and Technology University, Addis Ababa, Ethiopia}

\begin{abstract}

We discuss a newly established neutron monitor station installed at the ENTOTO Observatory Research Center outside of Addis Ababa, Ethiopia. This is a version of a mini-neutron monitor, {recently upgraded to detect individual neutrons and able to calculate the waiting time} distribution between neutron pulses down to $\sim 1 $ $\mu$s. From the waiting time distribution we define and calculate a new quantity, the correlation ratio, as the ratio of correlated to uncorrelated neutrons measured inside the monitor. We propose that this quantity can, in future, be used as a proxy for spectral index of atmospheric particles incident on the monitor and show that this quantity has a weak pressure dependence. We believe that future measurements from the ENTOTO mini-neutron monitor will contribute towards the understanding of cosmic ray acceleration and transport in the heliosphere.

\end{abstract}


\end{frontmatter}

\section{Introduction}

When primary cosmic rays interact with atmospheric molecules, showers of secondary atmospheric particles are created that can be registered, on ground-level, by neutron monitors \citep[NMs,][]{simpson2000}. The detectors integrate the energy-dependent atmospheric particle flux into a single count rate

\begin{equation}
    \mathcal{C}(P_c,t) = \sum_i \int_{P_c}^{\infty} j_i (P,t) Y_i (P,t, \ldots) dP,
\end{equation}

where $P_c$ is the cutoff rigidity (minimum rigidity particle that can reach the detector), $i$ represents the particle distribution under consideration (e.g. protons and heavier nuclei), and $Y_i (P,t, \ldots)$ is the so-called yield function that represents the response of the instrument on the unit flux of primary cosmic rays with rigidity $P$, including atmospheric and instrumental effects \citep[e.g.][]{clemdorman2000,mishevetal2020}.\\

NMs are increasingly used for several space weather applications, including the routine monitoring of the cosmic ray flux for dosimetric purposes, as well as environmental applications such as {sensing the soil moisture content \citep[e.g.][]{ZredaEtal2012}}. {With the expanding role of NMs, it is becoming increasingly important to fill the gaps in the international NM network \citep[e.g.][]{Mishev_Usoskin_2020}, especially over the African continent where NM stations are only located in Southern Africa.} In order to address this issue, we have recently installed a NM at the ENTOTO Observatory Research Center (EORC) outside of Addis Ababa, Ethiopia, jointly operated by the Ethiopian Space Science and Technology Institute (ESSTI) and the Centre for Space Research of the North-West University in South Africa.\\

First conceived as a mobile calibration monitor \citep[][]{krugeretal2008}, mini-neutron monitors (MNMs) have become an increasingly popular alternative to larger traditional (and much more expensive) NMs. The MNMs are cost-effective, relatively easy to transport and be installed, and features updated electronics. Due to their limited size MNMs, however, have a much lower count rate as traditional NMs but under the correct conditions (e.g. when installed at high altitudes), can still provide valuable measurements, even in a real-time setting \citep[][]{usoskin2015,stepan2015,heber2015}. Additionally, the updated electronics of the MNMs allow for the digitization of individually observed neutron pulses and the calculation of the waiting time distribution of the monitor down to $\sim 1$ $\mu$s \citep[][]{straussetal2020,Straussetal2021}. {In this work we consider three versions of the MNM. The first is a leaded version with a BF$_3$-filled proportional counter as installed at ENTOTO, the second a similar leaded version with a $^3$He gas-filled proportional counter, and the last version a BF$_3$-filled counter but with the lead producer removed.}\\

A long-term goal for the global NM network has been to inter-calibrate individual NM stations in order to derive a proxy for the spectral shape of the incident cosmic ray particles. While much progress regarding this has been made in recent years \citep[e.g.][]{mishevetal2020} there are examples where even very simplistic calculations, such as the ratio of two NMs at different cut-off rigidities, lead to spurious changes most likely related to environmental changes at the local NM stations \citep[][]{ruffoloetal2016}. This can be overcome by using two NMs, at the same station, which are sensitive to different rigidities, {usually a traditional NM with lead as well as a lead-free version \citep[][]{Stoker1985}} or by deriving a proxy for the incident spectra by examining the waiting time distribution of a single monitor \citep[][]{bieberetal2004}. Here we discuss how the waiting time distribution, as observed by the ENTOTO MNM can be used to derive such a proxy by calculating a new quantity, the {\it correlation ratio}, which we define as the ratio of correlated to uncorrelated {neutrons} measured by the detector.\\

\begin{figure*}[!t]
\begin{center}
\noindent\includegraphics[width=0.99\textwidth]{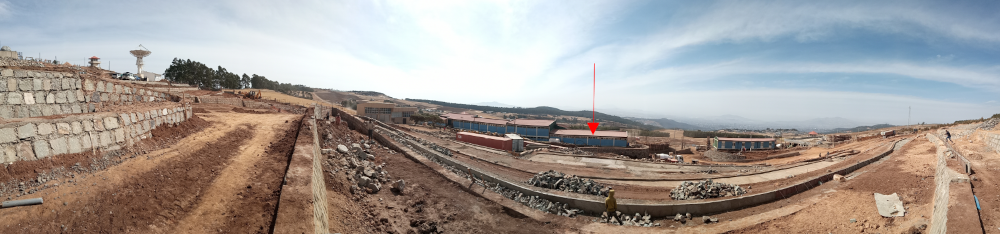}\\
\noindent\includegraphics[width=0.49\textwidth]{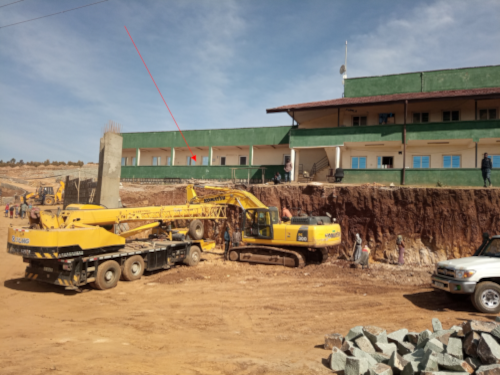}
\noindent\includegraphics[width=0.49\textwidth]{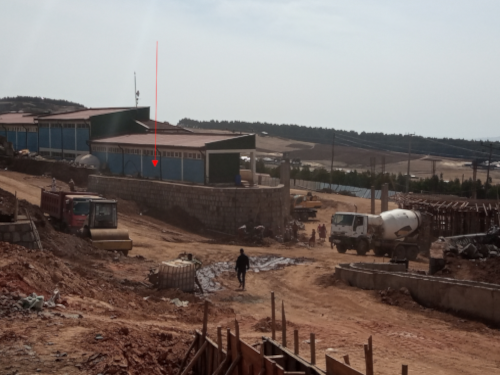}
\end{center}
\caption{Rear panoramic (top), front (bottom left), and back (bottom right) views of the ENTOTO Observatory Research Center currently undergoing major construction. The red arrows indicate the room where the MNM is currently installed.}
\label{Fig:pictures}
\end{figure*}

\begin{figure*}[!t]
\begin{center}
\noindent\includegraphics[width=0.99\textwidth]{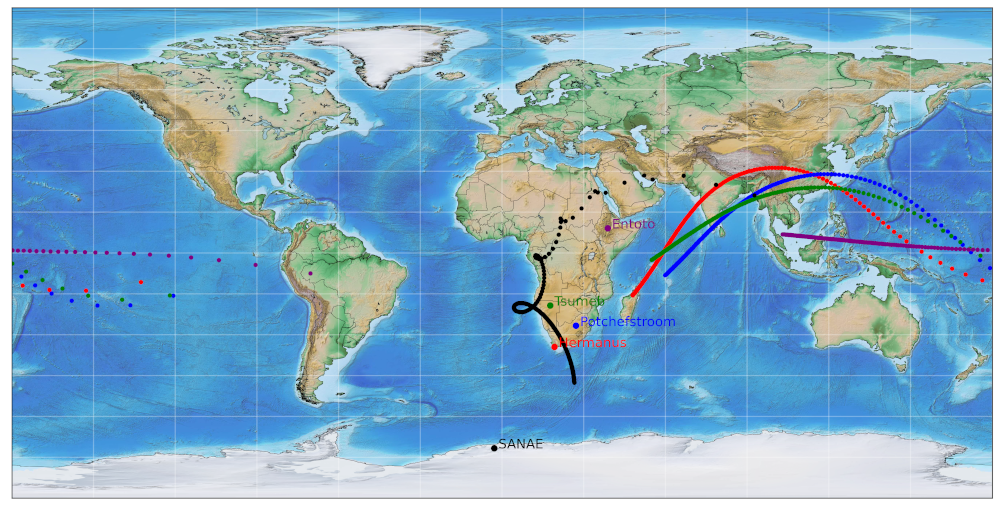}
\end{center}
\caption{The asymptotic cones of vertically incident primary particles reaching each monitor in the African neutron monitor network.}
\label{Fig:world_map}
\end{figure*}

\begin{figure*}[!t]
\begin{center}
\noindent\includegraphics[width=0.99\textwidth]{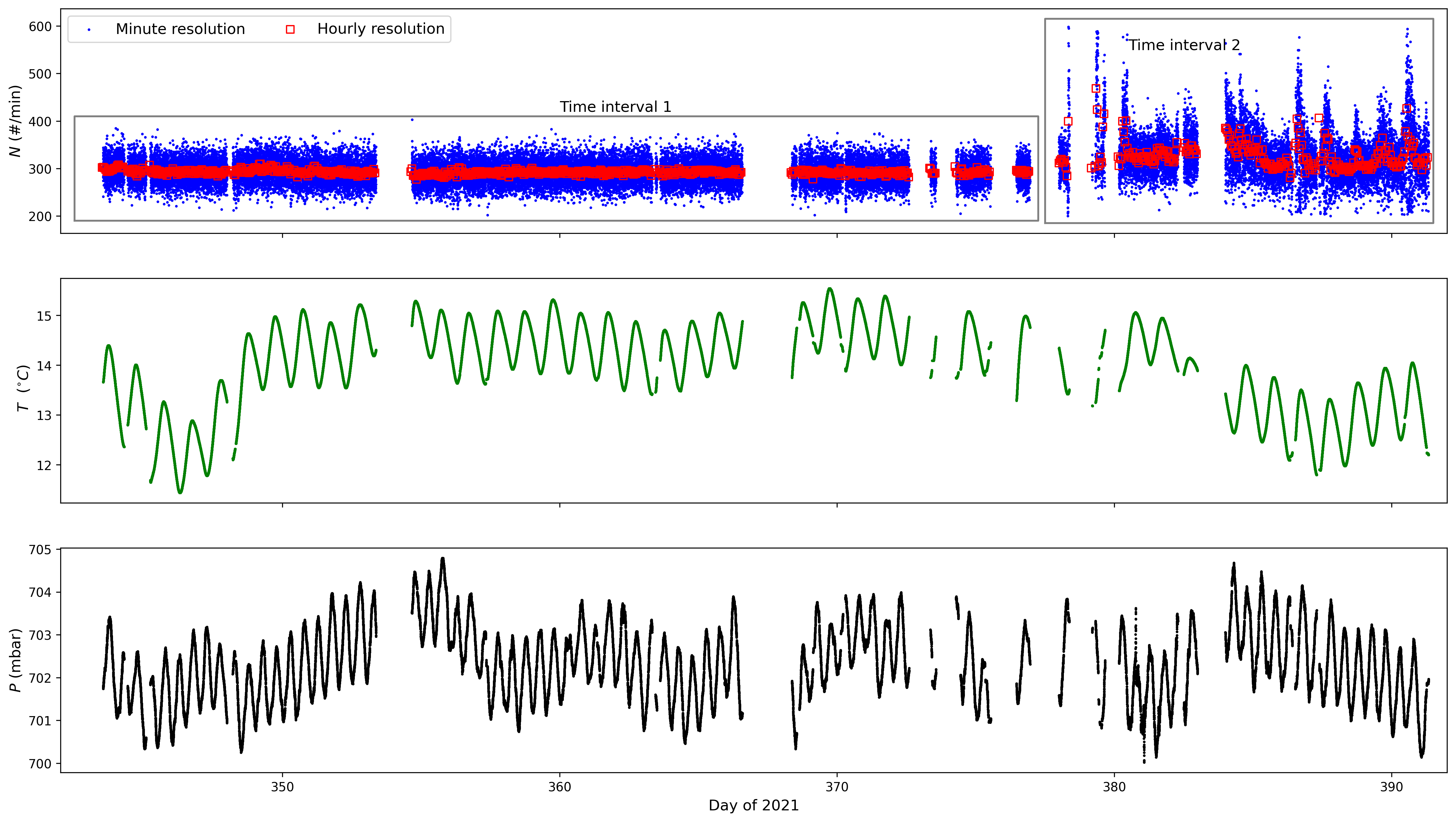}
\end{center}
\caption{The top panel shows the minute-averaged (blue) and hourly-averaged (red) count rate as a function of time. The middle and bottom panels show the measured temperature and pressure, respectively.}
\label{Fig:data_overview}
\end{figure*}

\begin{figure*}[!t]
\begin{center}
\noindent\includegraphics[width=0.99\textwidth]{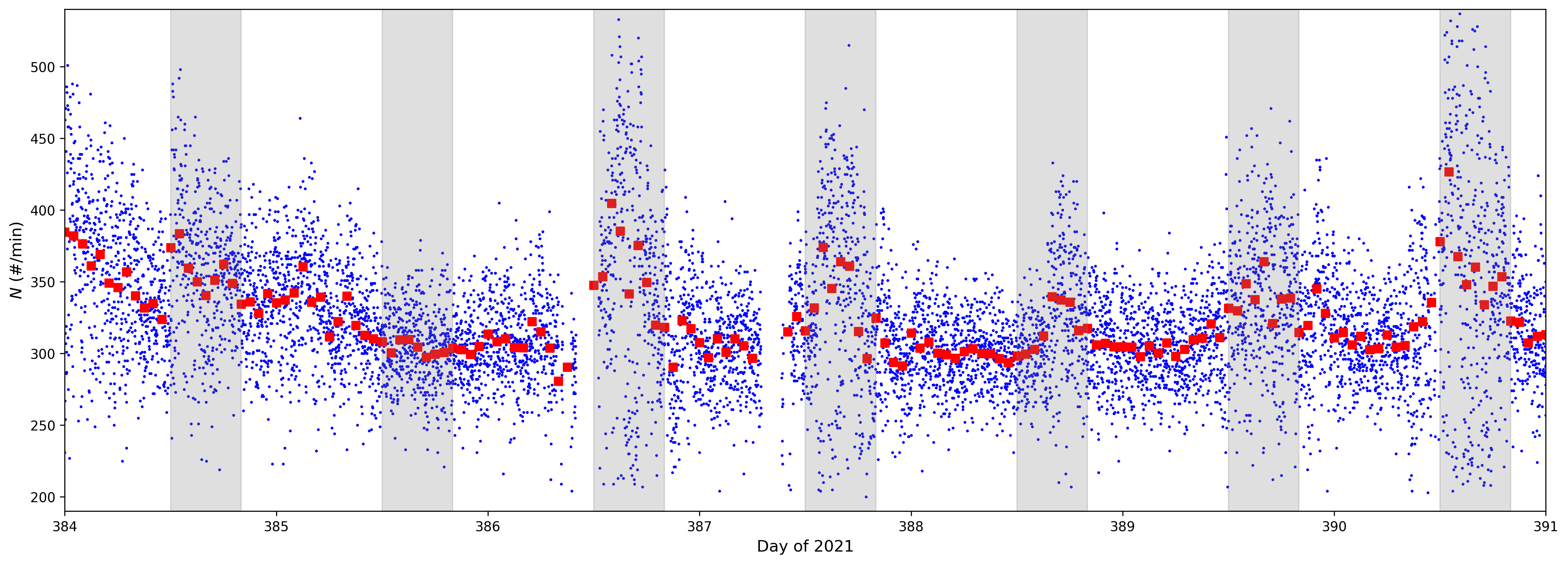}
\end{center}
\caption{A zoom in of time interval 2 as identified in Fig. \ref{Fig:data_overview}.}
\label{Fig:data_noise}
\end{figure*}

\section{The ENTOTO neutron monitor station}

\begin{table*}
\centering
\caption{Details of the newly established ENTOTO MNM, along with existing NM stations forming the African NM network.}
\label{Tab:entoto_details}
\begin{tabular}{|c|c|c|c|c|c|}
\hline
Station &Local time & Latitude & Longitude & Altitude & Cut-off rigidity\\
\hline \hline
ENTOTO, Ethiopia & UTC + 3 hr & 9$^{\circ}$ 6'  N & 38$^{\circ}$ 48'  E & 3175 m.a.s.l. &  16 GV \\
Tsumeb, Namibia & UTC + 2 hr & 19$^{\circ}$ 12' S & 17$^{\circ}$ 35' E  & 1240 m.a.s.l. & 9.2 GV \\
Potchefstroom, South Africa & UTC + 2 hr & 26$^{\circ}$ 41.9' S  & 27$^{\circ}$ 5.6' E  & 1351 m.a.s.l. & 7.2 GV \\
Hermanus, South Africa &  UTC + 2 hr & 34$^{\circ}$ 25.5' S & 19$^{\circ}$ 13.5' E & 26 m.a.s.l. & 4.9 GV \\
SANAE, Antarctica &  UTC & 71$^{\circ}$ 40' S  & 2$^{\circ}$ 51' W  & 856 m.a.s.l. & 0.8 GV \\
\hline
\end{tabular}
\end{table*}

A BF$_3$ version of the MNM, {recently reviewed in detail by \citet{straussetal2020}}, donated from the North-West University of South Africa to the Space Science and Applications Research and Development department of the Ethiopian Space Science and Technology Institute (ESSTI) of Ethiopia, was recently installed at the premises of the ENTOTO Observatory Research Center (EORC) located in the outskirts of Addis Ababa, the capital city of Ethiopia. Fig. \ref{Fig:pictures} shows recent pictures of this site which is currently undergoing major construction. The red arrows indicate the position where the MNM is currently installed. ENTOTO is {in} a mountainous landscape that reaches an altitude/elevation of 3200 m.a.s.l. In fact, the peak altitude of 3200 m.a.s.l is reached at the EORC premises. Since the name ENTOTO is associated with the vast and mountainous terrain that surrounds most of the northern and north-western outskirts of Addis Ababa, we have decided to refer to this MNM station as the ``ENTOTO Neutron Monitor", {abbreviated as ENTO}. Fig. \ref{Fig:world_map} shows the asymptotic cones of acceptance for the African NM network using the model of \citet{smartetal2000}. The highest rigidity particle considered is 30 GV (the data point closest to the station), decreasing linearly until the cut-off rigidity is reached (data point furthest away from the station). The ENTOTO MNM, being installed very close to the magnetic equator, primarily measures particles incident along the Earth's equatorial regions. Details of the ENTOTO NM station is summarized in Tab. \ref{Tab:entoto_details}, along with existing NM stations forming the African NM network. \\

\subsection{Unprocessed results}

The raw data from the ENTOTO MNM is shown in Fig. \ref{Fig:data_overview}. The top panel shows the minute-averaged (blue) and hourly-averaged (red) count rate as a function of time with two time intervals identified. These are discussed further in later sections. The average (uncorrected for pressure changes) count rate of the monitor is $\sim 290$/min. The middle and bottom panels show the measured temperature and pressure, respectively. The data gaps is caused by power outages at the installation site.\\

\subsection{Construction noise}

During early 2022, major construction unfortunately started at the ENTOTO observatory site. Mechanical vibration and possible electrical interference from these activities are believed to contribute to the higher levels of noise present during time interval 2, identified in Fig. \ref{Fig:data_overview}. The count rate data from this time interval is shown again in Fig. \ref{Fig:data_noise}, {where the shaded regions indicate times where the majority of the construction occurs: 9:00 - 17:00 local time}. Unfortunately this additional noise component cannot be corrected for and the data from time interval 2 is not used for further analysis in the rest of the work.\\

\section{The Neutron monitor waiting time distribution}


\subsection{Origin of neutron monitor multiplicities}

{For either of the $^3$He or BF$_3$ filling gases, an electrical pulse is produced via a neutron capture process \citep{knoll12010}, i.e.}

\begin{eqnarray}
^{10}\mathrm{B} + \mathrm{n} & \rightarrow & \left\{ \begin{array}{ccc}
     ^{7}\mathrm{Li} + \alpha + 2.78 \, \mathrm{MeV} \\
     ^{7}\mathrm{Li}^{*} + \alpha + 2.30 \, \mathrm{MeV} \end{array} \right. 
\end{eqnarray}

or

\begin{equation}
    ^{3}\mathrm{He} + \mathrm{n} \rightarrow \,  ^{3}\mathrm{H} +\mathrm{p} + 0.76 \, \mathrm{MeV}
\end{equation}

that ionizes a small portion of the filling gas. The pulse-height measured in the NM is therefore independent of energy of the incident atmospheric particles. Low- and high-energy atmospheric hadrons, however, interact with the lead producer in different ways. Low energy protons and neutrons interact with the {monitor}, and thus the proportional gas, in a random fashion and their detected pulse show a stochastic nature with subsequent pulses unrelated to each other. A high energy neutron or proton, on the other hand, can interact with the lead producer to form a number of low energy {\it evaporation neutrons} \citep[e.g.][]{bieberetal2004}. These neutrons are observed as a number of pulses over a short period of time, showing a high level of temporal correlation. This is usually expressed in terms of the waiting time distribution (i.e. the time between subsequent pulses observed in the monitor) as the level of multiplicity (i.e. the number of pulses observed in a certain time increment). By thus looking at the waiting time distribution observed by the monitor (examples are shown in the next section) these two components can be distinguished; a low multiplicity component at long waiting times and a high multiplicity, highly correlated, component at short waiting times. By examining the ratio of these components, which we refer to as the {\it correlation ratio}, some information of the incident particle spectrum can, in principle, be reconstructed \citep[see e.g.][]{ruffoloetal2016,mangeard2016,Bangliengetal2020}.   \\

\subsection{An illustrative example: Lead vs. lead-free neutron monitors}

\begin{figure*}[!t]
\centering
\includegraphics[width=0.99\textwidth]{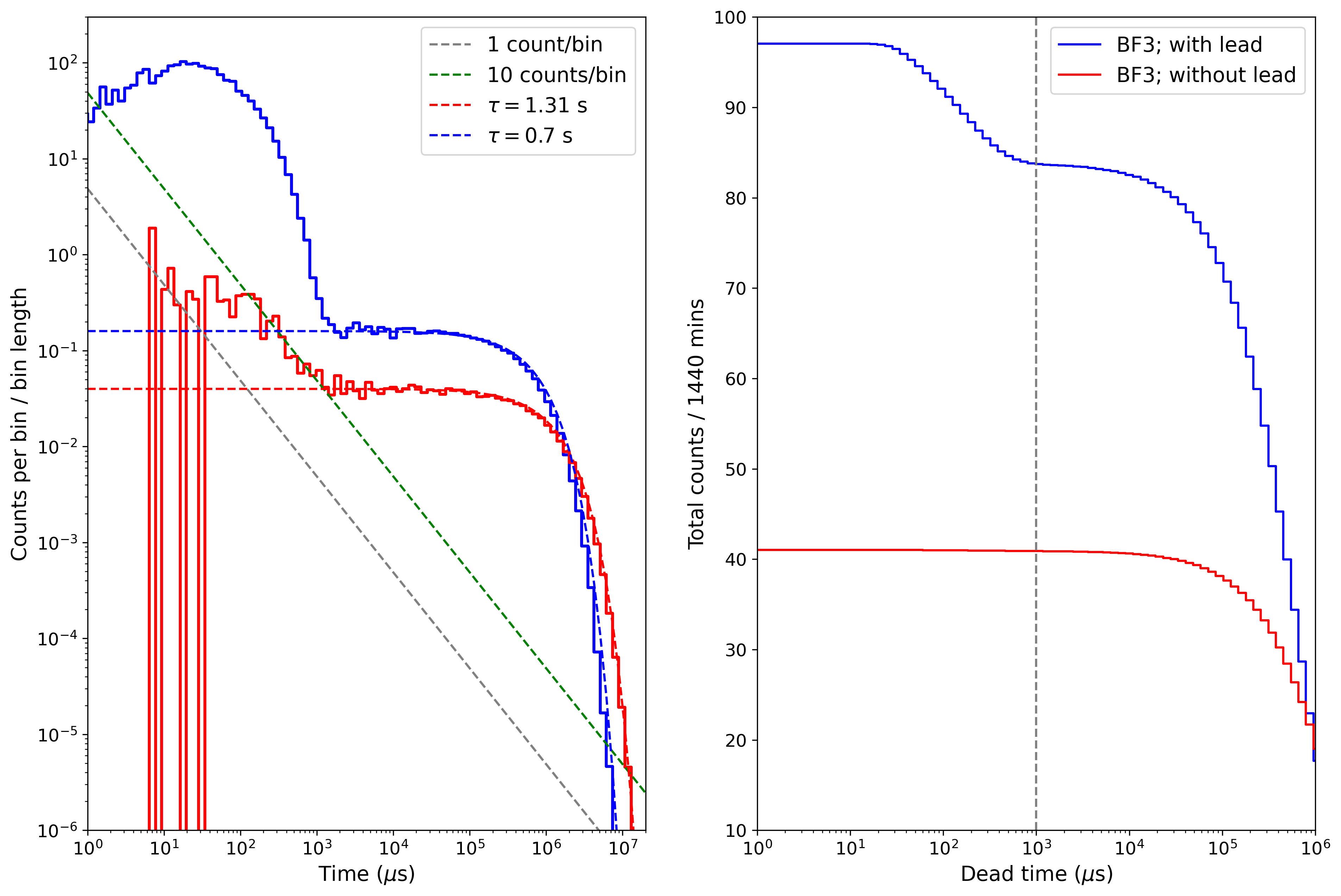}
\caption{The waiting time distribution (left panel) and integrated waiting time distribution (right panel) for two identical BF$_3$-filled mini-NMs; one operated with lead (blue curves) and one without (red curves). In the left panel, the 1 and 10 counts limits are shown (gray and green dashed curves), while the low multiplicity portions of the distributions are fitted with {exponential} distributions (blue and red dashed curves). The vertical dashed line on the right panel indicates the approximate change from high to low multiplicities observed in the detector with lead at $\sim 1$ms.}
\label{fig:Multi_BF3}
\end{figure*}

To illustrate how the waiting time distribution is formed by different neutron populations, we performed the following test: Two identical BF$_3$ filled MNMs were {placed} next to each other in the physics laboratory in Potchefstroom. The MNMs were operated for a number of days to confirm that the electronics were also identical, with identical count rates (within statistical variations) obtained from both monitors. The lead in one monitor was then removed and both MNMs operated for a day. The resulting waiting time distributions (simply calculated as the probability distribution of the time elapsed between successive counts in the detector) are shown in the left panel of Fig. \ref{fig:Multi_BF3}, where the low multiplicity (long waiting time) distribution is fitted with the following {exponential} distribution 

\begin{equation}
    \mathcal{D}(\Delta t) = \mathcal{D}_0  \exp \left( \frac{\Delta t - \, ^{0}\Delta t}{\tau} \right),
    \label{Eq:posson}
\end{equation}

where $\mathcal{D}_0$, $^{0}\Delta t$, and $\tau$ are fit constants and $\Delta t$ is the waiting time between successive counts \citep[see also the discussion by][]{ruffoloetal2016}.\\

Clearly, the monitor with the lead producer (blue curve in Fig. \ref{fig:Multi_BF3}) shows the high multiplicity (high level of temporal correlation) distribution which is absent in the lead-free monitor (red curve in Fig. \ref{fig:Multi_BF3}). However, the low multiplicity distributions are also significantly different with a higher count rate observed, even at longer waiting times, for the monitor with lead (roughly a factor of 2). A higher count rate also changes the low multiplicity {exponential} distribution (see Eq. \ref{Eq:posson}), with the lead monitor well described with a $\tau$ parameter roughly twice that of the lead-free monitor. The low multiplicity part (at longer waiting times) are detected neutrons that are therefore consistent with particles interacting randomly with the detector with the resulting counts unrelated. \\

In the right panel of Fig. \ref{fig:Multi_BF3}, we calculated, using measurements presented in the left panel, the total number of counts measured by these detectors as a function of simulated dead time of the detector, $\Delta T$,

\begin{equation}
\label{Eq:integrate_total}
   \mathcal{C}(\Delta T) = \int_{\Delta T}^{\infty} \mathcal{D}(\Delta t) \Delta t,
\end{equation}

which is essentially an integration of the waiting time distribution over all possible waiting times. The vertical dashed line at 1 ms indicates the changeover from the low to the high multiplicity distribution. {The leaded monitor has an average count rate of $\sim 95$/min. From the integrated multiplicity distribution we estimate that the low multiplicity (long waiting time) distribution of the monitor with lead has an average count rate of $\sim$80/min while the high multiplicity (short waiting time) distribution contributes at a rate of $\sim$15/min. This suggests that, for the leaded monitor, $\sim$20\% of the measured neutrons are high multiplicity neutrons produced in the lead producer by high energy protons and/or neutrons, while the remaining $\sim$80\% are low multiplicity neutrons produced either in the lead producer or in the atmosphere. The monitor without lead gives an average count rate of $\sim$40/min which are presumably due to neutrons produced in the atmosphere. This suggests that, for the BF$_3$ version of the mini-NM located in Potchefstroom, about half of low multiplicity neutrons measured by the leaded monitor are already formed in the atmosphere, contributing with a rate of $\sim$40/min towards the total count rate.} It is also instructive to remember that $\sim 90\%$ of the atmospheric particles detected by a typical NM are atmospheric neutrons \citep[][]{mangeard2016}. Also note that these estimates will be different for different NM designs and will depend on the local environment and incident particle spectrum.\\

\section{Measuring and quantifying the waiting time distribution}

\subsection{Correlated and uncorrelated count rates}
\label{Sec:distributions}

\begin{figure*}[!t]
\begin{center}
\noindent\includegraphics[width=0.49\textwidth]{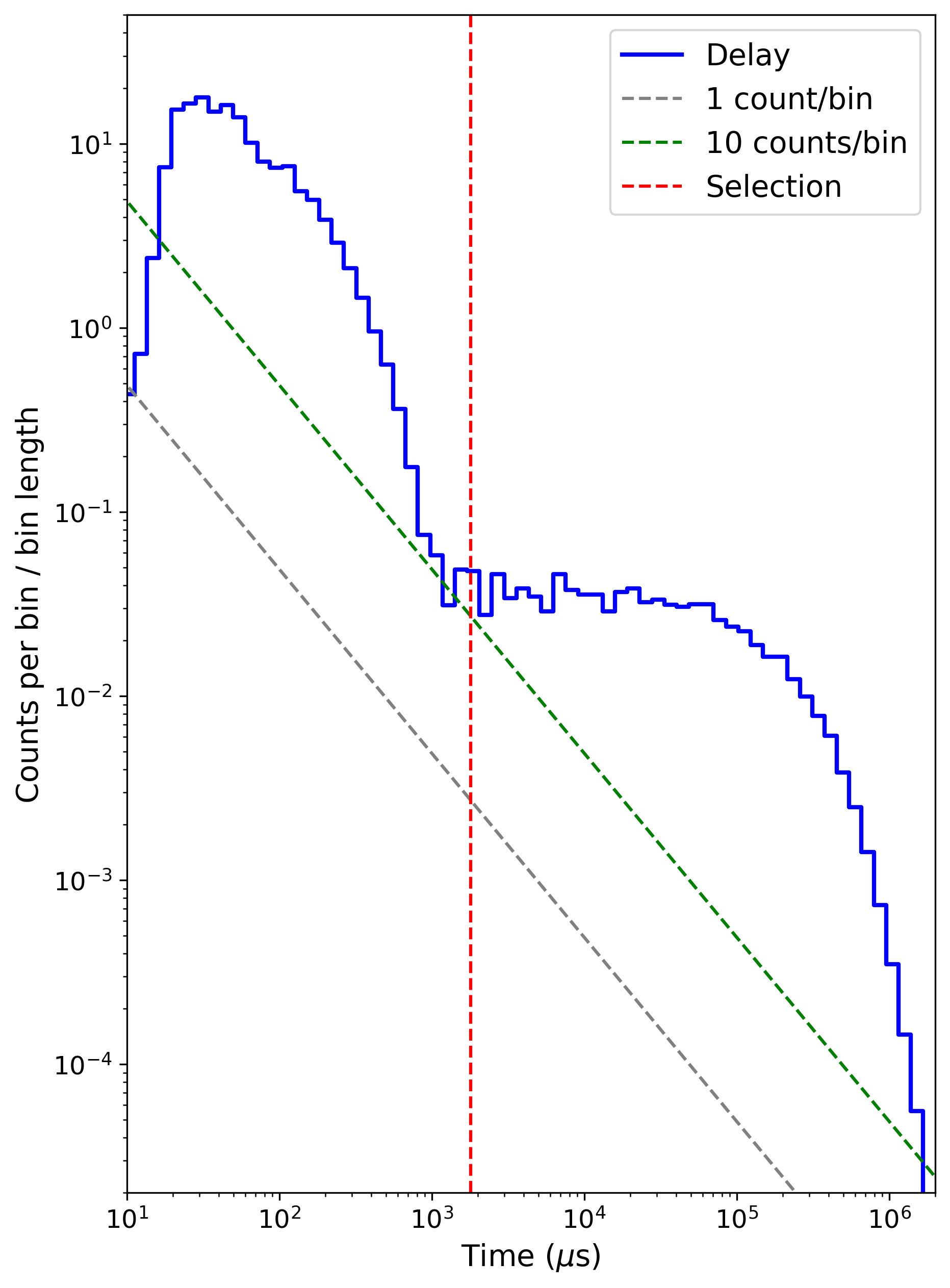}
\noindent\includegraphics[width=0.49\textwidth]{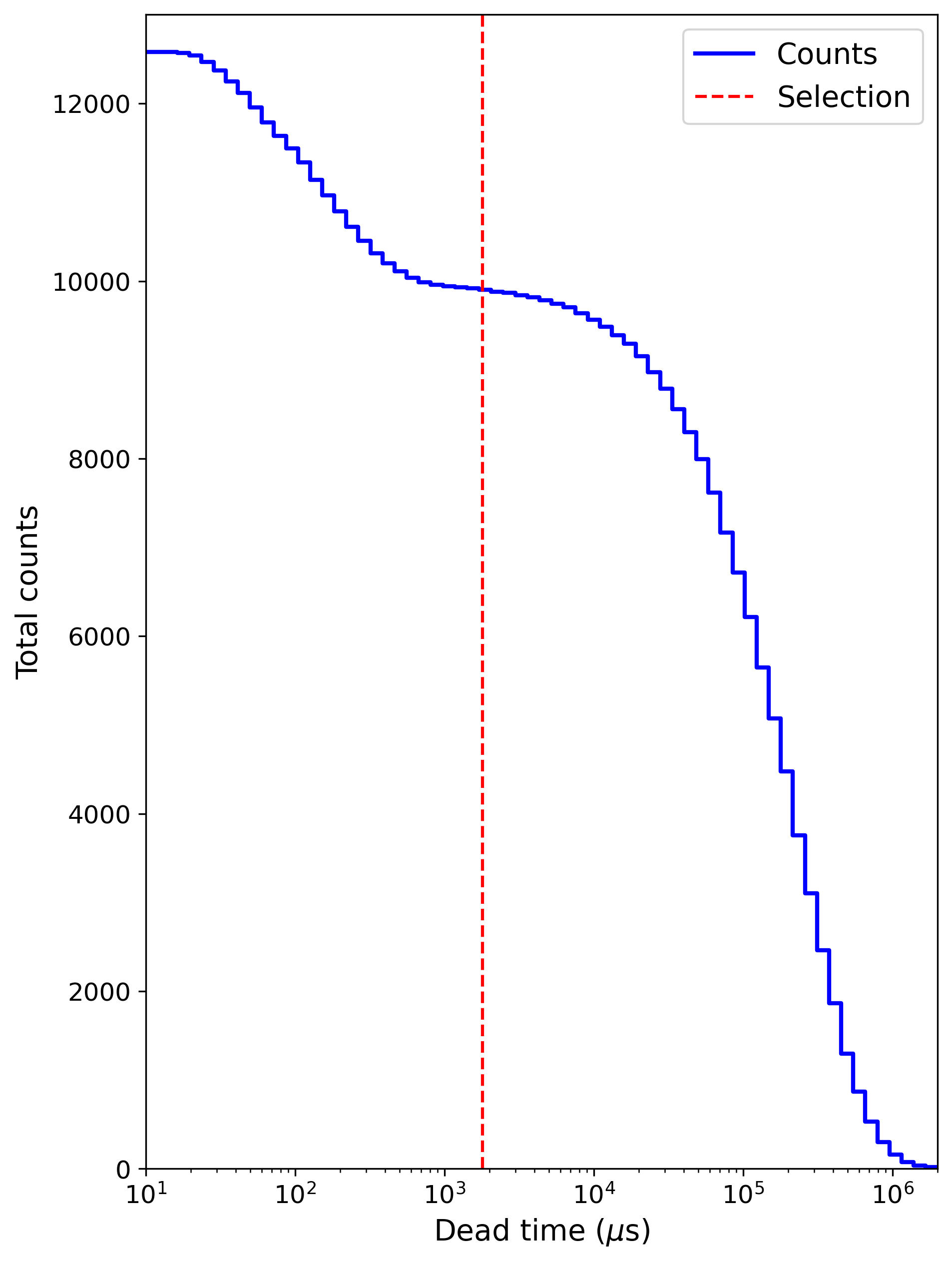}
\end{center}
\caption{Similar to Fig. \ref{fig:Multi_BF3}, but now calculate using a hour's measurements of the ENTOTO MNM.}
\label{Fig:entoto_distribution}
\end{figure*}

\begin{figure*}[!t]
\begin{center}
\noindent\includegraphics[width=0.99\textwidth]{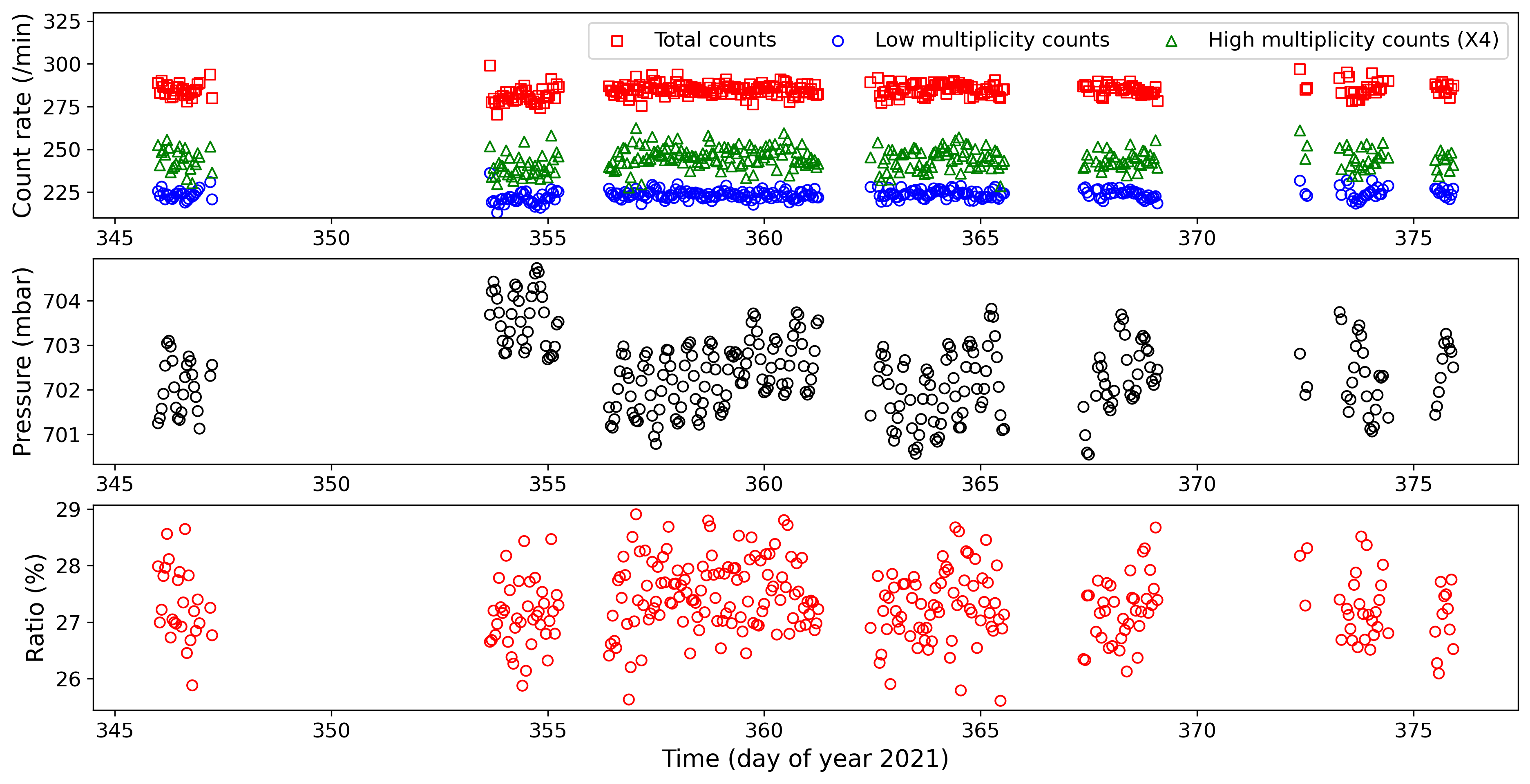}
\end{center}
\caption{The top panel shows hourly averaged count rates for the low multiplicity (blue), high multiplicity (green, multiplied by a factor of 4), and the total distributions. The middle panel shows the hourly averaged pressure and the bottom panel the ratio of the high multiplicity to low multiplicity count rates, the so-called correlation ratio. }
\label{Fig:entoto_different_countrates}
\end{figure*}

The waiting time distribution as measured by the ENTOTO MNM is shown in Fig. \ref{Fig:entoto_distribution}, with a similar format as Fig. \ref{fig:Multi_BF3}. As expected, the uncorrelated low multiplicity (long waiting times, $>2$ ms) and correlated high multiplicity (short waiting times, $<2$ ms) distributions can be identified. The vertical red dashed line in Fig. \ref{Fig:entoto_distribution}, at $\Delta T' = 2$ ms, indicates the assumed change in distribution and used for the rest of the work to separate the two distributions. \\

We now calculate the hourly waiting time distribution, similar to {Fig.} \ref{Fig:entoto_distribution}, for each hour in time interval 1 (as defined in Fig. \ref{Fig:data_overview}). From these distributions, we calculate the hourly count rate of the high multiplicity component as

\begin{equation}
   \mathcal{C}_{\mathrm{high}}(t,\Delta T') = \int_{0}^{\Delta T'} \mathcal{D}(t,\Delta t) \Delta t,
\end{equation}

the count rate of the low multiplicity component as

\begin{equation}
   \mathcal{C}_{\mathrm{low}}(t,\Delta T') = \int_{\Delta T'}^{\infty} \mathcal{D}(t,\Delta t) \Delta t,
\end{equation}

and the total count rate as

\begin{equation}
   \mathcal{C}_{\mathrm{total}}(t) =  \mathcal{C}_{\mathrm{low}}(t,\Delta T') +   \mathcal{C}_{\mathrm{high}}(t,\Delta T') = \int_{0}^{\infty} \mathcal{D}(t,\Delta t) \Delta t.
\end{equation}

These different count rates are shown as a function of time in the top panel of Fig. \ref{Fig:entoto_different_countrates} with the hourly averaged pressure in the middle panel. {Note that for these calculations we only use complete hourly intervals with a full 60 minutes of measurements. The data is therefore much more sparse than the minute-averaged measurements shown in Fig. \ref{Fig:data_overview} which contains a large number of data gaps.} In the bottom panel we calculate the so-called {\it correlation ratio}, as

\begin{equation}
\mathcal{R}(t,\Delta T') = \frac{\mathcal{C}_{\mathrm{high}}(t,\Delta T')}{\mathcal{C}_{\mathrm{low}}(t,\Delta T')} \times 100\%,
\end{equation}

which we expect to be a proxy for the spectral index of the incident atmospheric particles. The data presented here gives a ratio of $\sim 27\%$ {when using a value of $\Delta T' =$ 2ms}. As this is essentially the ratio of correlated to uncorrelated counts in the monitor, we expect the ratio to increase (more correlated, high multiplicity counts) as the spectrum of incident particles become harder (more high energy particles). We test for any pressure dependence of the different count rates, and the ratio thereof, in the next section.

\begin{figure*}[!t]
\begin{center}
\noindent\includegraphics[width=0.49\textwidth]{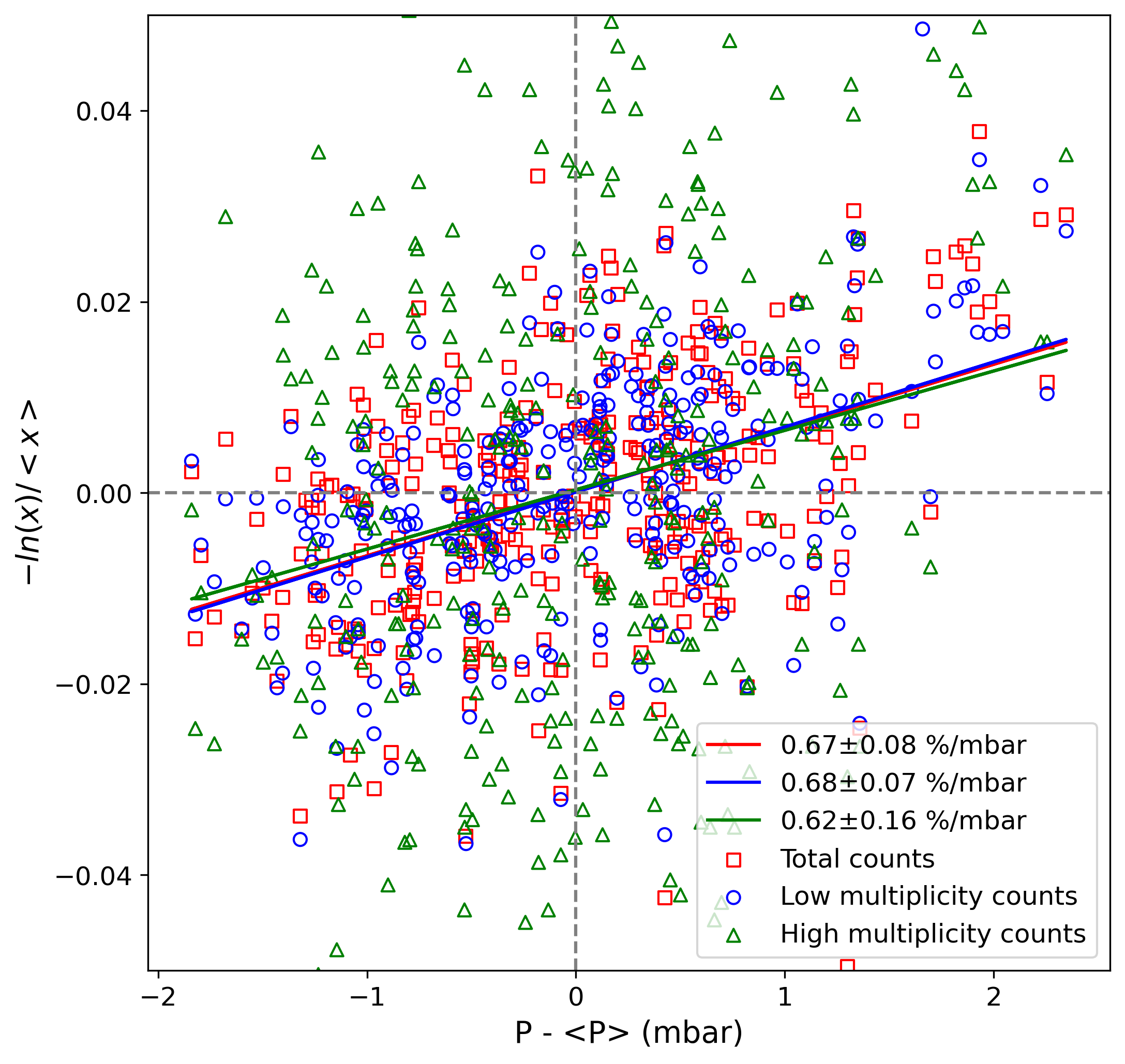}
\noindent\includegraphics[width=0.49\textwidth]{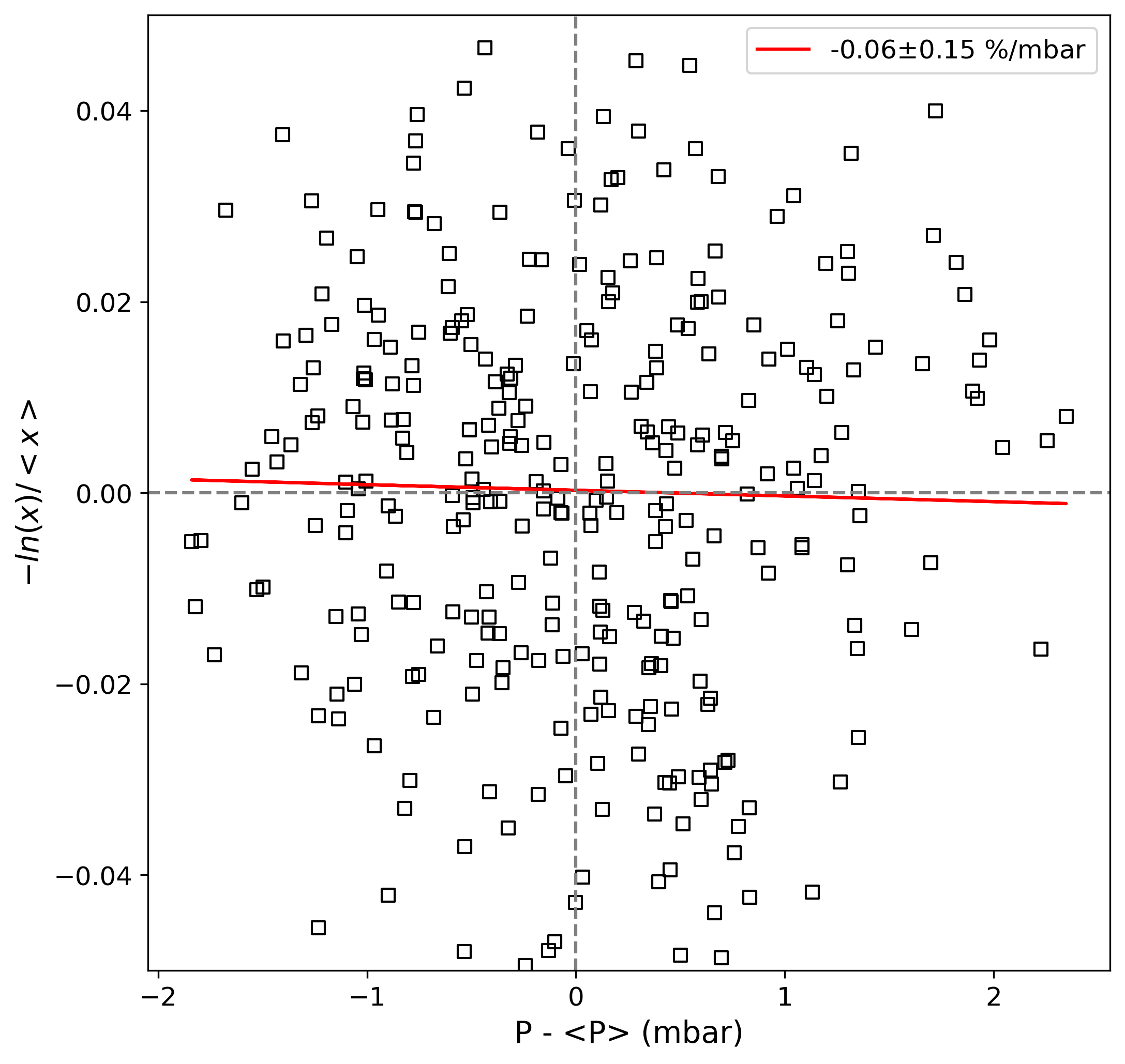}
\end{center}
\caption{The left panel shows the barometric correction applied separately for the 3 different count rates identified in Sec. \ref{Sec:distributions}, while the right panel shows the same calculation for the correlation ratio.}
\label{Fig:entoto_pressure_corrections}
\end{figure*}

\subsection{Pressure corrections}

Assuming the standard exponential correlation between the count rate and atmospheric pressure, the hourly count rate is expected to vary with varying atmospheric pressure, $P$, as

\begin{equation}
    \mathcal{C}_{\mathrm{x}} = \langle \mathcal{C}_{\mathrm{x}} \rangle \exp \left( -\beta_{\mathrm{x}} \Delta P \right),
\end{equation}

where $x$ refers to either the low, high, or total count rate, $\Delta P = P - \langle P \rangle$, and $\langle \cdot \rangle$ indicates long-term averaging. The barometric coefficients, $\beta_{\mathrm{x}}$, for the different count rates are calculated in the left panel of Fig. \ref{Fig:entoto_pressure_corrections} as the gradient of a linear regression, with resulting values included in the legend where the colour of the regresion line corresponds to that of the data points. As the high multiplicity counts are produced by higher energy atmospheric particles, we expect the corresponding barometric coefficient to be smaller than of the low multiplicity component, based on the work of \citet{2021AdSpR_torado_bueno}. While there might be an indication of this, the different coefficients are the same within the (relatively large) fit errors.\\

The correlation ratio is expected to have a pressure dependence of the form,

\begin{equation}
    \mathcal{R} = \frac{\mathcal{C}_{\mathrm{high}}}{\mathcal{C}_{\mathrm{low}}} = \frac{\langle \mathcal{C}_{\mathrm{high}} \rangle}{\langle \mathcal{C}_{\mathrm{low}} \rangle} \exp \left[ -\left(\beta_{\mathrm{high}} - \beta_{\mathrm{low}} \right) \Delta P \right],
\end{equation}

with an effective pressure coefficient of $\beta = \beta_{\mathrm{high}} - \beta_{\mathrm{low}}$. The pressure dependence of $\mathcal{R}$ is therefore expected to be weaker than that of the individual count rates. This is confirmed in the right panel of Fig. \ref{Fig:entoto_pressure_corrections} where we apply a similar pressure correction to the correlation ratio. Within the fit errors we find the correlation ratio to be independent of pressure changes.\\

\begin{figure*}[!t]
\begin{center}
\noindent\includegraphics[width=0.44\textwidth]{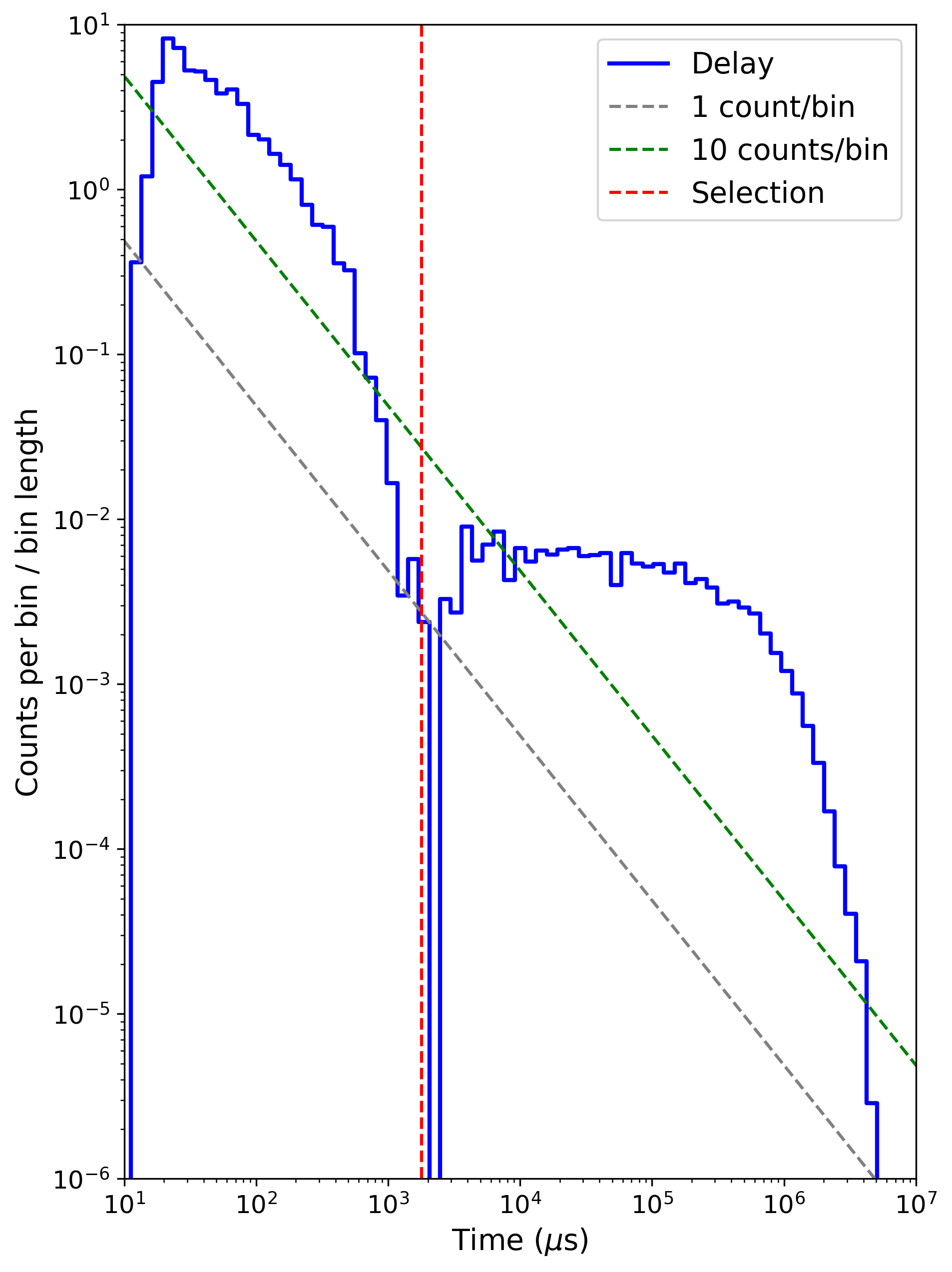}
\noindent\includegraphics[width=0.42\textwidth]{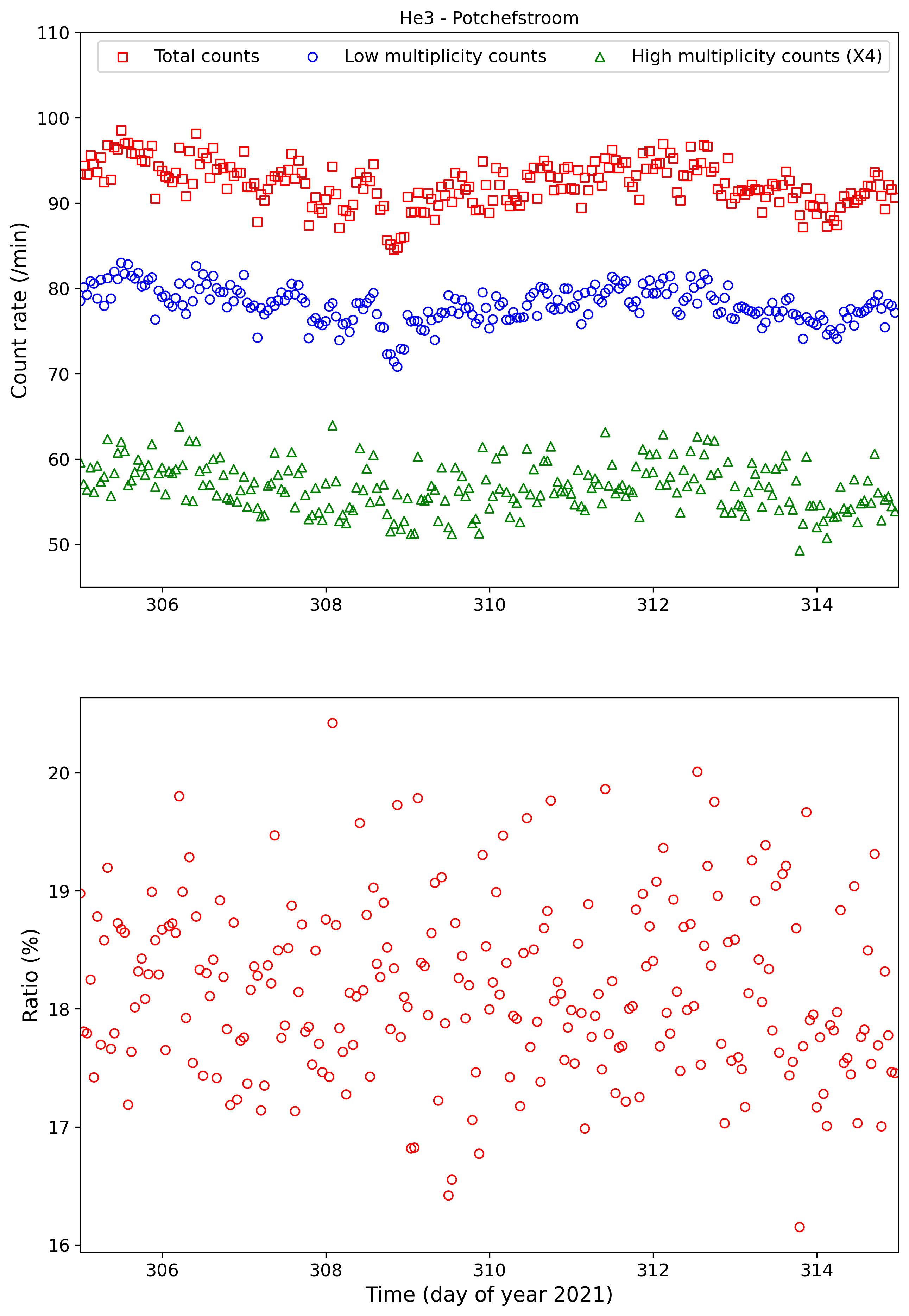}\\
\noindent\includegraphics[width=0.44\textwidth]{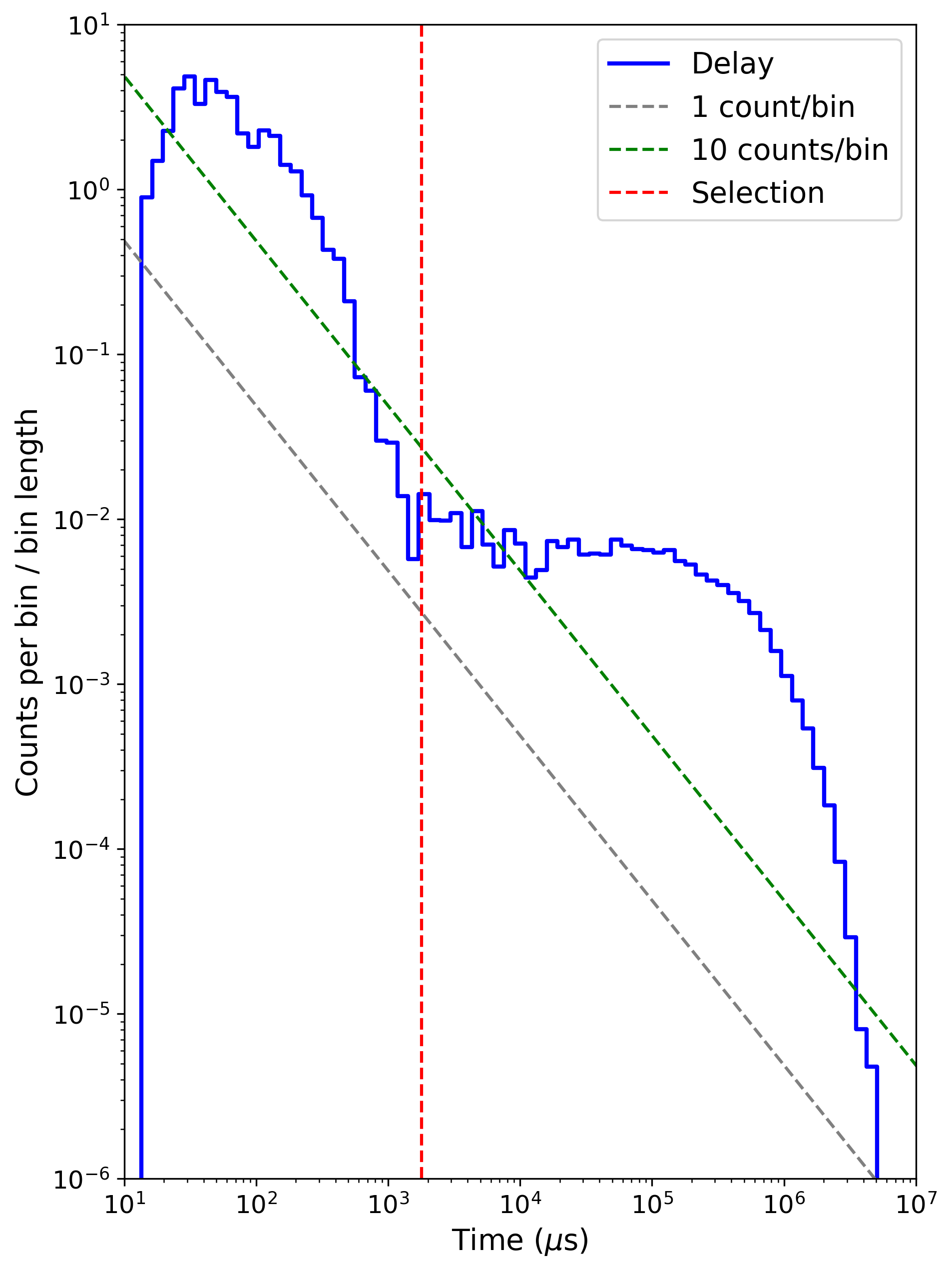}
\noindent\includegraphics[width=0.42\textwidth]{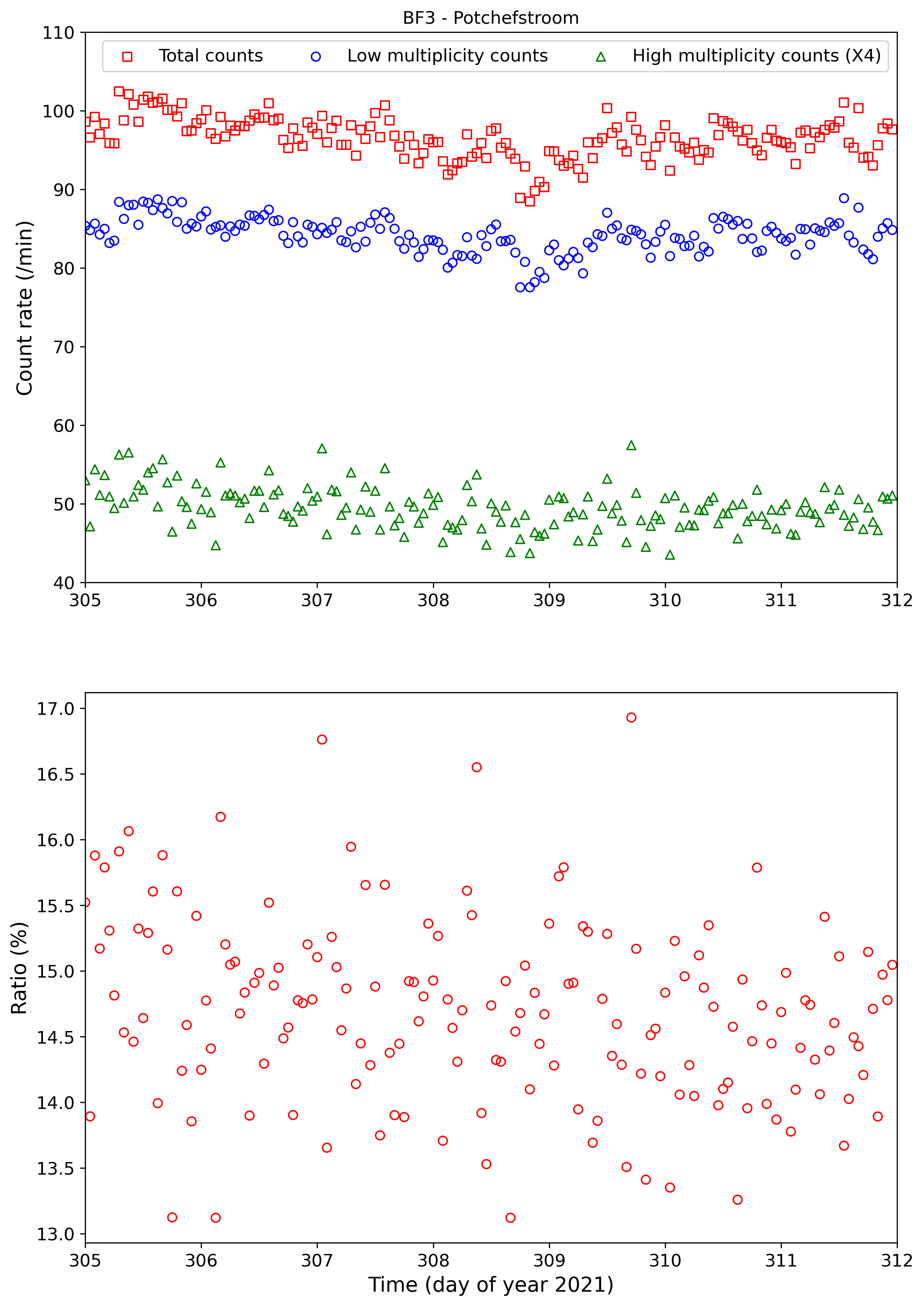}
\end{center}
\caption{Similar to Figs. \ref{Fig:entoto_distribution} and \ref{Fig:entoto_different_countrates}, but now for a $^3$He (top panels) and a BF$_3$ (bottom panels) MNM housed in the laboratory in Potchefstroom.}
\label{Fig:he3lab}
\end{figure*}

\subsection{Comparison with other monitors}

We end the analysis presented here with a comparison between the ENTOTO waiting time distribution and the time distributions measured by a $^3$He and BF$_3$ leaded MNM housed in the laboratory in Potchefstroom. The results for these MNMs are shown in Fig. \ref{Fig:he3lab}. Of course, the waiting time distributions show similarities with that of ENTOTO, with the change between high and low multiplicity distributions again occurring close to $\sim 2$ ms. The right panels of the figure show the behaviour of the different count rates, and the correlation ratio, as a function of time. For the $^3$He tube, the ratio is $\sim 18\%$, and for the BF$_3$ tube it is $\sim 14\%$. It should be kept in mind that the $^3$He version of the MNM has a thicker lead producer (due to a thinner tube) than the BF$_3$ version which might explain the additional $\sim 3\%$ of high-multiplicity neutrons detected.\\

\section{Summary, conclusions, and outlook}
\label{Sec:sumary}

We have discussed a newly established MNM station installed at the ENTOTO Observatory Research Center outside of Addis Ababa, Ethiopia. Preliminary data from the ENTOTO MNM is shown and discussed, with a special focus on characterising the waiting time distribution.  \\

Using the waiting time distribution, as observed by this new station, we calculated the ratio of correlated to uncorrelated counts as a proxy for the spectral index of the incident atmospheric particles. While similar proxies were used in the past, e.g. the leader fraction of \citet{ruffoloetal2016}, our {\it correlation ratio}, as defined here, is a natural characterization of the waiting time distribution as measured by our MNMs. We hope to also calculate the correlation ratio for other NMs in future, as well track how this ratio changes during temporal events such as ground level enhancement and Forbush decreases.\\

We calculated the pressure dependence of different count rates of the ENTOTO MNM, and found very similar values for both the correlated and uncorrelated count rates. This suggests that the pressure dependence of correlation ratio will be rather weak and, for typical MNM count rates, probably negligible as the uncertainty in the measurements is dominated by the statistical errors associated with low count rates. \\

The ENTOTO MNM, at a cut-off rigidity of $\sim 16$ GV, has a current correlation ratio of $\sim 27\%$. We also calculated this quantity for two MNMs (with a BF$_3$ and $^3$He tube) currently undergoing testing at the laboratory in Potchefstroom. These MNMs recorded ratios of $\sim 14\%$ and $\sim 18\%$, respectively. Given the fact that Potchefstroom has an approximate cut-off rigidity of only $\sim 7$ GV, the higher ENTOTO correlation ratio can be explained by the fact that the ENTOTO MNM measures higher rigidity particles, resulting in a higher ratio of correlated neutrons being detected inside the monitor, partially validating our proposal that this ratio will serve as a proxy for the spectral shape of the incident atmospheric particles. We will continue to investigate the use of the correlation ratio for this purpose, especially during transient phenomena such as Forbush decreases, and report such results in due course. \\

\section*{Acknowledgement}

Figures prepared with Matplotlib \citep{Matplotlib-2007} and certain calculations done with NumPy \citep{harrisetal2020}. This work is based on the research supported in part by the National Research Foundation of South Africa (NRF grant numbers: 119424, 120847, and 120345). Similarly the authors would like to acknowledge the Ethiopian Space Science and Technology Institute (ESSTI) for hosting, maintaining, and monitoring the instrument. Opinions expressed and conclusions arrived at are those of the authors and are not necessarily to be attributed to either the NRF or the ESSTI. {We acknowledge support provided by the North-West University (NWU) Instrument Making Department.}


\newpage 

\bibliographystyle{model5-names}
\biboptions{authoryear}
\bibliography{mybibfile}

\end{document}